\documentclass[11pt,twoside]{article}

\usepackage{asp2014}

\aspSuppressVolSlug
\resetcounters

\begin{document}

\title{Pre-feasibility Study of Astronomical Data Archive Systems Powered by Public Cloud Computing and Hadoop Hive}

\author{Satoshi Eguchi$^1$
\affil{$^1$Department of Applied Physics, Faculty of Science, Fukuoka University, Fukuoka, Fukuoka Prefecture, Japan; \email{satoshieguchi@fukuoka-u.ac.jp}}}

\paperauthor{Eguchi~Satoshi}{satoshieguchi@fukuoka-u.ac.jp}{}{Fukuoka University}{Department of Applied Physics}{Fukuoka}{Fukuoka Prefecture}{814-0180}{Japan}

\begin{abstract}
The size of astronomical observational data is increasing yearly.
For example, while Atacama Large Millimeter/submillimeter Array is
expected to generate 200 TB raw data every year, Large Synoptic
Survey Telescope is estimated to produce 15 TB raw data every night.
Since the increasing rate of computing is much lower than that of
astronomical data, to provide high performance computing (HPC)
resources together with scientific data will be common in the next decade.
However, the installation and maintenance costs of a HPC system can
be burdensome for the provider.
I note public cloud computing for an alternative way to get
sufficient computing resources inexpensively.
I build Hadoop and Hive clusters by utilizing a virtual private server (VPS)
service and Amazon Elastic MapReduce (EMR), and measure their performances.
The VPS cluster behaves differently day by day, while the EMR clusters
are relatively stable.
Since partitioning is essential for Hive, several partitioning algorithms
are evaluated.
In this paper, I report the results of the benchmarks
and the performance optimizations in cloud computing environment.
\end{abstract}

\section{Introduction}

The size of astronomical observational data has been increasing year by year.
For example, while Atacama Large Millimeter/submillimeter Array (ALMA) is
expected to generate 200 TB raw data every year \citep{2014SPIE.9149E..02S},
Large Synoptic Survey Telescope (LSST) will produce 15 TB raw data every night
\footnote{\url{https://www.lsst.org/sites/default/files/docs/sciencebook/SB_Whole.pdf}}.
Though high performance computing (HPC) systems are required to process such big data,
they are so expensive and occupy large physical spaces, that is,
the systems are hardly available except for some large project teams.
Hence sharing HPC resources together with huge observational data over the Internet
will be common in the next decade.
On the other hand, alternative ways to handle relative large astronomical data
at low cost will be in demand.
From this aspect, I note public cloud computing and see if it is applicable
to astronomical data.
In this paper, I focus on Hadoop, which is an open-source software framework
for distribute file system and data processing, and Hive, which is a SQL-like
distributed database system running on Hadoop clusters;
Hadoop is designed to run on a cluster consisting of a large number of cheap
standard PCs, and hardware failures are automatically recovered by re-execute
of the failed jobs on other nodes.
I report the results of the benchmarks and performance optimizations.

\section{Challenges in Hadoop and Hive}

Hadoop and Hive are designed to process a set of moderate large files in parallel;
a massive number of very small files exhaust memory resources to manage
their metadata on HDFS, which is the native distributed file system of Hadoop.
On the other hand, a small number of huge files are inefficient since they lead to
very frequent data transportation between nodes over the network and intensive
file I/O, and remarkably reduce a degree of parallelism.

Different from wide-spread standard RDBMSs, Hive does not manage datasets by indexes.
Instead, datasets can be organized by ``partitions'', which correspond to directories
on HDFS and seem to be one of keys of a table.
Files to be read or processed in a Hive query are narrowed down by specifying
the values of the partitions.

There are two ways to get Hadoop clusters in cloud computing:
Virtual Private Server (VPS) and Infrastructure as a Service (IaaS).
In the former case, options of configurations of virtual hardware are quite limited,
and a Virtual Machine (VM) can be created only from a VM image with a pre-installed
operating system provided by the service provider;
users cannot instantiate a VM from their own images.
In addition, to build a Hadoop cluster is time consuming since the users need to
install and setup the software packages by hand on each VM instance.
On the other hand, in the latter case, a wide variety of hardware configurations
are available, and users can also create a VM instance from their own images.
Furthermore, we can construct a Hadoop and Hive cluster by one command
in case of Amazon Elastic MapReduce (EMR).

\section{Strategies of bechmarking}

To evaluate applicability of VPS and IaaS to astronomical data,
I develop a simple benchmark program in Java, which connects to Hive via
the JDBC driver.
Data files for 2MASS Catalog Server Kit\footnote{\url{http://www.ir.isas.jaxa.jp/~cyamauch/2masskit/}}
are used as test data.
I add two columns below to the dataset:
\texttt{healpix\_id}, a HEALPix ID with $N_{\rm side}^{\rm pixel} = 2^{16}$
(fixed) of each source position,
and \texttt{healpix\_partition}, a HEALPix ID of a source
with $N_{\rm side}^{\rm partition} = 2^{3}, 2^{4}, \cdots$.\footnote{The number of pixels are given by $N_{\rm pixel} = 12 N_{\rm side}^{12}$.}
Firstly, querying positions and search radii ($5^{\prime \prime}$--$5^{\prime}$)
are randomly selected with uniform distributions by Mersenne Twister.
At this stage, the random seed is fixed at a certain value.
Secondly, the range of \texttt{healpix\_partition} is calculated
by the HEALPix library implemented in Java.
Thirdly, the angular distances are computed for all the rows in the given
\texttt{healpix\_partition} range based on \citet{2011PASP..123.1324Y}.
Lastly, the average magnitudes of J, H, K-bands are calculated
with the built-in function \texttt{AVG()} in Hive for the sources
within the given search radius.
These procedures essentially emulate an use case to cut out desired data cubes
from $\ge$3-dimensional high resolution all-sky images.

\section{Results}

\subsection{VPS}

A Hadoop and Hive cluster consisting of 1 NameNode (master node)
and 7 DataNodes (slave nodes) by utilizing ``Small Plan'' provided
by GMO CLOUD K.K., which is a Japanese IT company.
Each node has 4 CPU cores, 4 GB RAM, and 200 GB HDD, costing about
\$230 annually.
Pure Apache Hadoop and Hive distributions are used,
that is, Tez is not installed.
The construction of a partition tree is performed on a workstation
and transported to the cluster via the SSH protocol due to
the small memory size of the nodes and a restriction of our firewall.
Figure~\ref{fig-vps-day-by-day} represents the distributions of
searching time with $N_{\rm size}^{\rm partition} = 2^{3}$
measured on different days, suggesting that the performance
of the cluster differs day by day.

\articlefigure[width=0.5\linewidth]{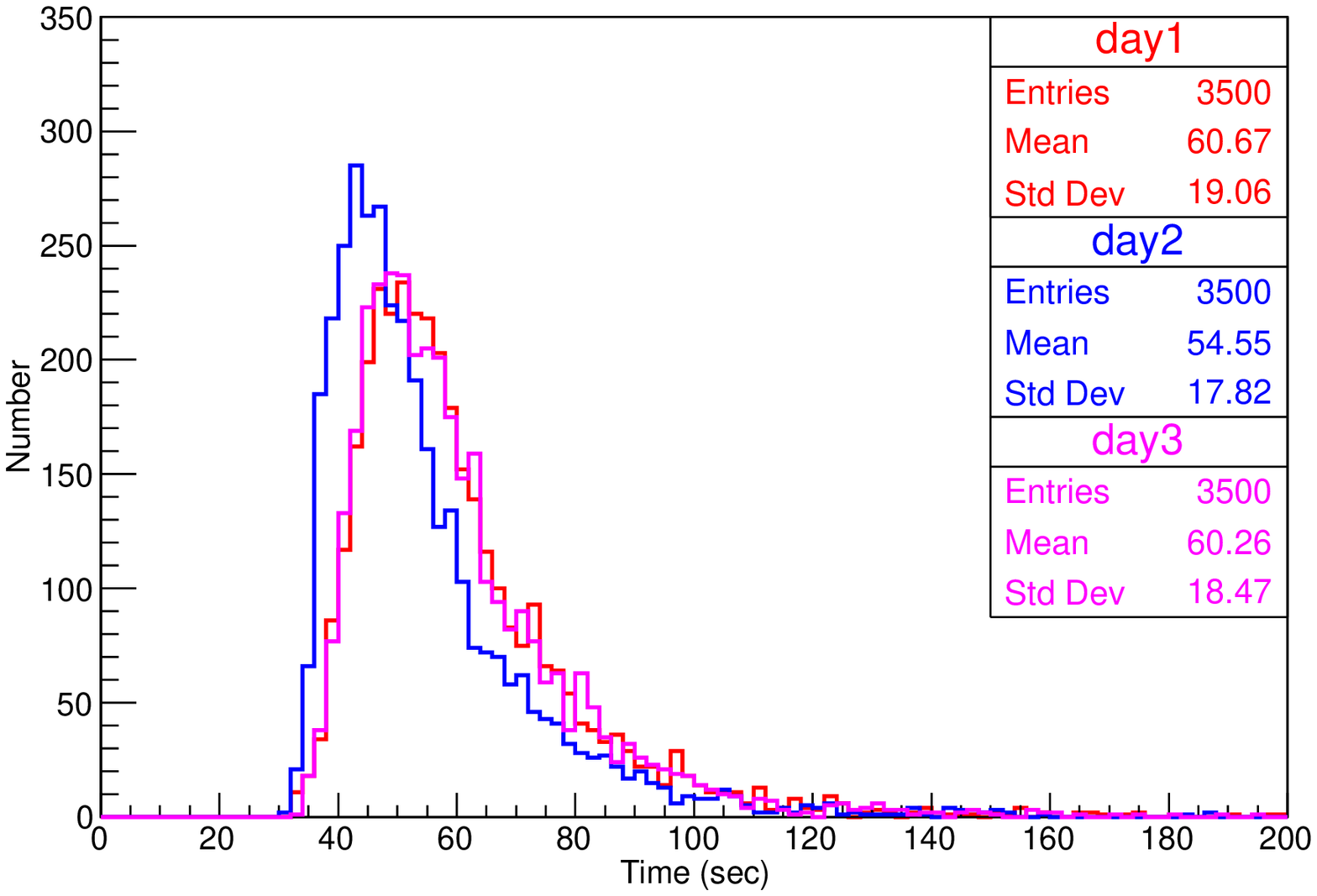}{fig-vps-day-by-day}{The searching time distributions of the VPS cluster measured on different days.}

\subsection{Amazon EMR as an IaaS solution}

As an IaaS solution, I use Amazon EMR.
Clusters with 1 NameNode and 3 DataNodes are created with \texttt{m3.xlarge} instances,
each of which has 4 CPU cores and 15 GB RAM and costs \$0.385 an hour, every time
when a new set of the benchmark is started.
Physical files of the database are stored on Amazon S3, which is a cloud storage
service provided by Amazon and can be accessed seamlessly from EMR instances,
since files on HDFS are lost when the cluster is terminated.
Note that the VM instances and database files locate in the Tokyo region.
At this stage, Tez is enabled.

The left and right of Figure~\ref{fig-emr} represent the distributions of searching time
with $N_{\rm side}^{\rm partition} = 2^{3}$ executed on different cluster instances,
and the dependence of mean searching time on $N_{\rm side}^{\rm partition}$, respectively;
the former suggests that no difference between the instances are observed,
and the latter shows that mean searching time decreases as $N_{\rm side}^{\rm partition}$
increases.
$N_{\rm side}^{\rm partition} \ge 2^{7}$ are not measured due to insufficient memory.
Further investigation reveals that a typical query is distributed to only 3 nodes
since the range of \texttt{healpix\_partition} is $\le 3$ for $N_{\rm side}^{\rm partition}
\le 2^{6}$.
However, a wide range of \texttt{healpix\_partition} is found to make
the time required to schedule at the initialization stage much longer.

\articlefiguretwo{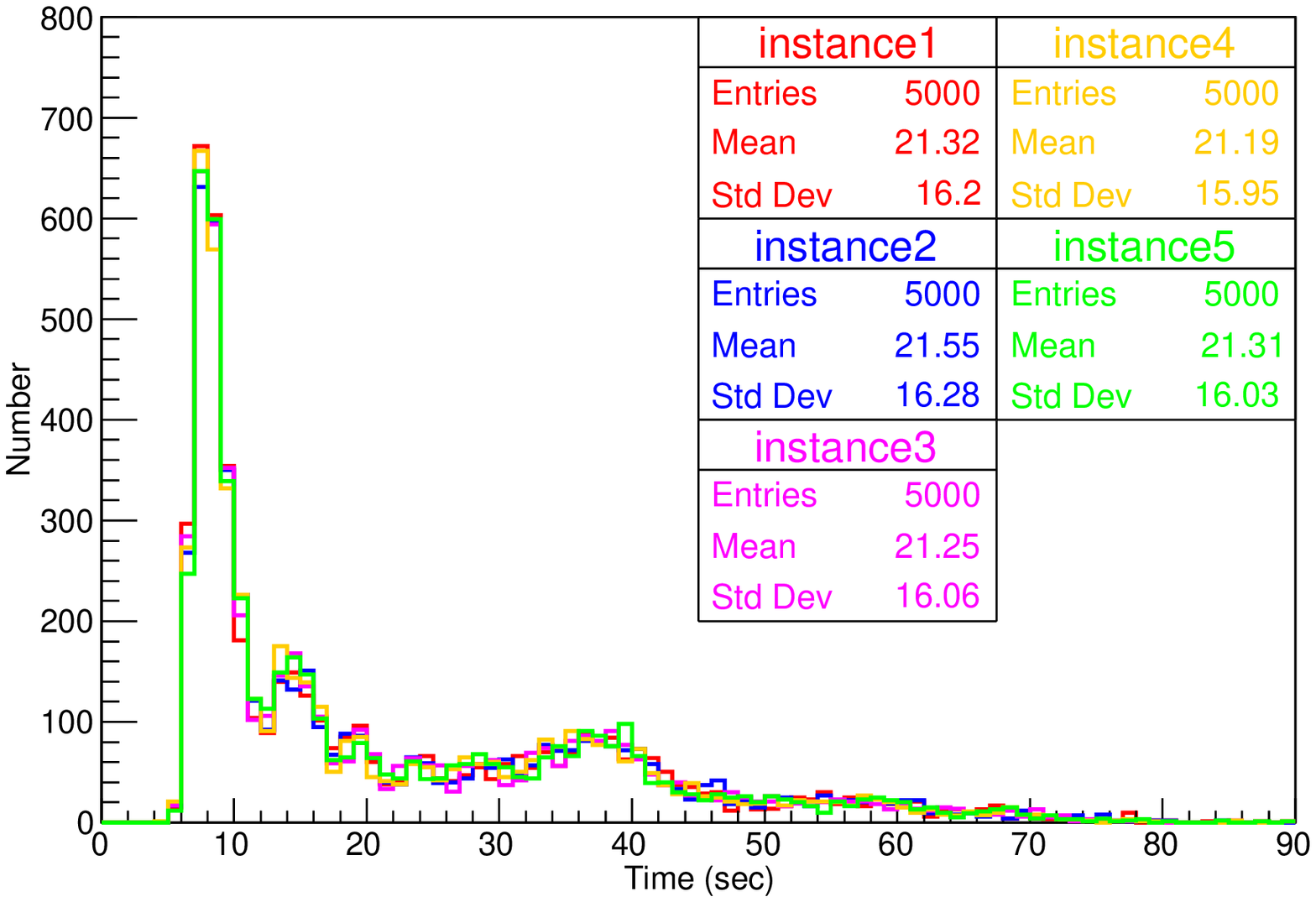}{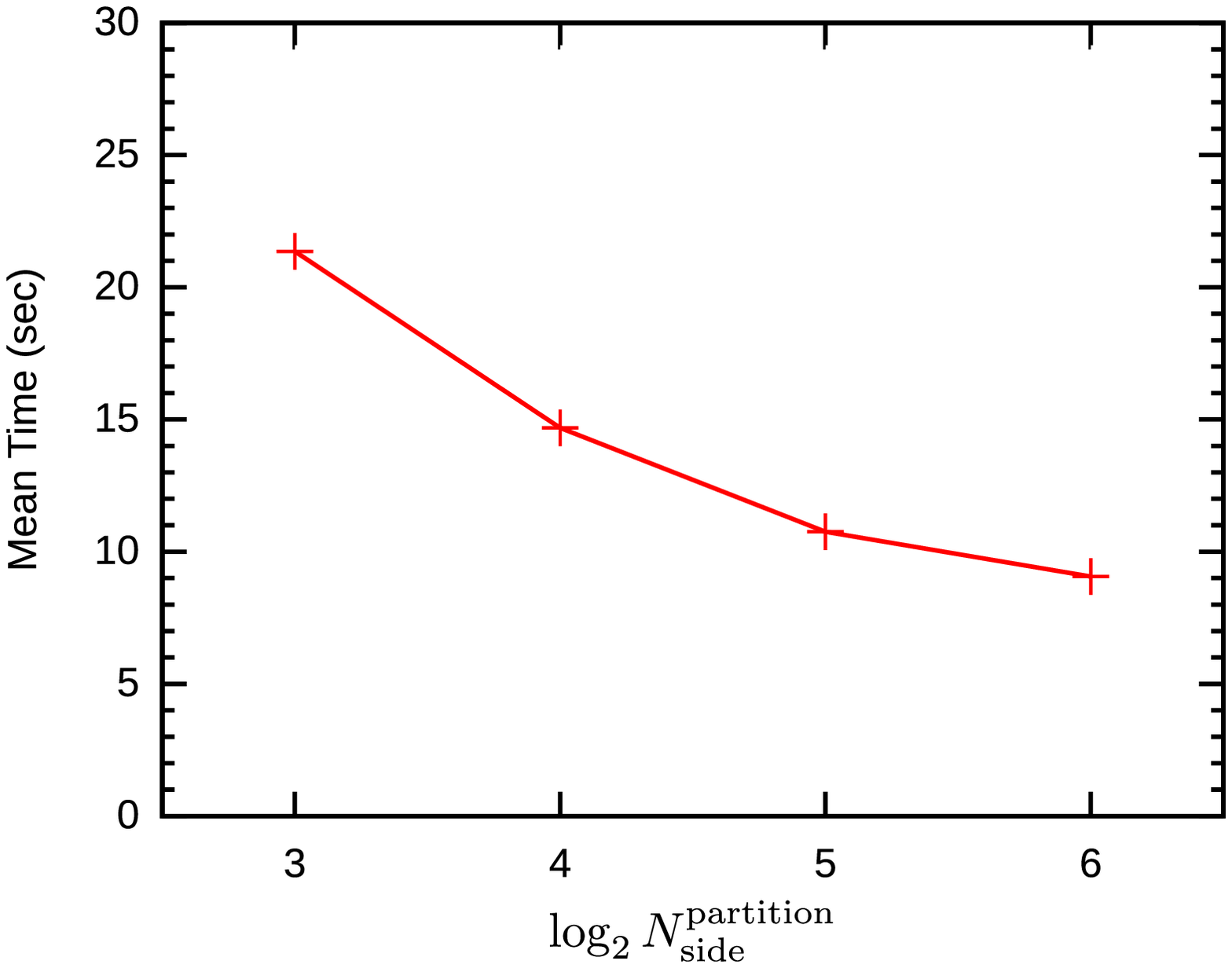}{fig-emr}{The distributions of searching time on different instances (left) and the dependence of the mean searching time on $N_{\rm side}^{\rm partition}$ (right).}

\subsection{Nested partitions on EMR}

The introduction of \texttt{healpix\_id} modulo 16 ($\equiv$\texttt{healpix\_mod})
into the $N_{\rm side}^{\rm partitions} = 2^{3}$ case and partitioning
the dataset into the pair of (\texttt{healpix\_partition}, \texttt{healpix\_mode})
reduce an effective file size of one partition to that in the $N_{\rm side}^{\rm partition} = 2^{5}$
case.
Hence this approach is expected to make the searching time
in the $N_{\rm side}^{\rm partition} = 2^{3}$ case same as
that in the $N_{\rm side}^{\rm partition} = 2^{5}$ case.
However, an application of this method to EMR does not change the searching time at all.

\section{Future work}

\begin{itemize}
 \checklistitemize
 \item Identification of parameters controlling the degree of parallelism
 \item Checking if a larger number of partitions are possible
\end{itemize}

\acknowledgements

This work is supported by JSPS KAKENHI Grant Numbers JP15K17501.

\bibliography{P3-4}  

\end{document}